\documentclass[aps,prd,twocolumn,nofootinbib,floatfix,preprintnumbers]{revtex4}
\usepackage{subfigure}
\usepackage{graphicx}
\usepackage{dcolumn}
\usepackage{epsfig}
\usepackage{amsmath}
\usepackage{amsfonts}
\usepackage{amssymb}
\usepackage{color}
\usepackage{hyperref}

\newcommand{\be}{\begin{equation}}
\newcommand{\ee}{\end{equation}}
\newcommand{\bea}{\begin{eqnarray}}
\newcommand{\eea}{\end{eqnarray}}
\newcommand{\nn}{\nonumber}

\begin{document}

\title{The Linear Point Standard Ruler for galaxy survey data: validation with mock catalogues}

\author{Stefano Anselmi}
\email{stefano.anselmi@iap.fr}
\affiliation{LUTH, UMR 8102 CNRS, Observatoire de Paris, PSL Research University, Universit\'e Paris Diderot, 92190 Meudon -- France}
\affiliation{Institut d'Astrophysique de Paris, CNRS UMR 7095 and UPMC, 98bis, bd Arago, F-75014 Paris -- France}

\author{Pier-Stefano Corasaniti}
\affiliation{LUTH, UMR 8102 CNRS, Observatoire de Paris, PSL Research University, Universit\'e Paris Diderot, 92190 Meudon -- France}

\author{Glenn D.~Starkman}
\affiliation{Department of Physics/CERCA/Institute for the Science of Origins, Case Western Reserve University, Cleveland, OH 44106-7079 -- USA}

\author{Ravi K.~Sheth}
\affiliation{Center for Particle Cosmology, University of Pennsylvania, 209 S. 33rd St., Philadelphia, PA 19104 -- USA}
\affiliation{The Abdus Salam International Center for Theoretical Physics, Strada Costiera, 11, Trieste 34151 -- Italy}

\author{Idit Zehavi}
\affiliation{Department of Astronomy, Case Western Reserve University, Cleveland, OH 44106-7079 -- USA}

\date{\today}

\begin{abstract}

Due to late time non-linearities, the location of the acoustic peak in the two-point galaxy correlation function is a redshift-dependent quantity, thus it cannot be simply employed as a cosmological standard ruler. This has motivated the recent proposal of a novel ruler, also located in the Baryon Acoustic Oscillation range of scales of the correlation function, dubbed the {\it linear point}. Unlike the peak, it is insensitive at the $0.5\%$ level to many of the non-linear effects that distort the clustering correlation function and shift the peak. However, this is not enough to make the linear point a useful standard ruler. In addition, we require a model-independent method to estimate its value from real data, avoiding the need to deploy a poorly known non-linear model of the correlation function. In this manuscript, we precisely validate a procedure for model-independent estimation of the linear point.  We also identify the optimal set-up to estimate the linear point from the correlation function using galaxy catalogs. The methodology developed here is of general validity, and can be applied to any galaxy correlation-function data. As a working example, we apply this procedure to the LOWZ and CMASS galaxy samples of the Twelfth Data Release (DR12) of the Baryon Oscillation Spectroscopic Survey (BOSS), for which the estimates of cosmic distances using the linear point have been presented in  Anselmi et al. (2017) \cite{2017arXiv170301275A}.
\end{abstract}


\maketitle

\newcommand{\ste}[1]{\textcolor{red}{\textbf{\small[Ste: #1]}}}


\section{Introduction}
\label{intro}
 
Baryon Acoustic Oscillations (BAO) in the late-time matter power spectrum result from primeval acoustic waves propagating in the coupled baryon-photon plasma before decoupling \cite{1970ApJ...162..815P, 1966JETP...22..241S, 1968ApJ...151..459S}. These manifest as a peak in the two-point correlation function (CF) of galaxies, located at the scale of the sound horizon at the so-called drag epoch, when the acoustic waves stop freely propagating through the plasma. This provides a natural comoving standard ruler to constrain the cosmic expansion history \cite{1998ApJ...504L..57E,Bassett:2009mm,2005ApJ...633..560E}. 

Ideally, one would estimate cosmic distances by measuring the location of the BAO peak directly from CF data, without the need to model the processes that shape the CF. Unfortunately, on BAO scales the late-time distribution of matter is sensitive to the non-linear dynamics of matter's gravitational clustering. Several studies, using both high-precision cosmological simulations and analytic models, have shown that non-linearities distort the BAO pattern: smearing the BAO peak, lowering its amplitude and shifting its position \cite{2008PhRvD..77d3525S, 2007ApJ...664..660E,Matsubara:2007wj,Crocce:2007dt}. Therefore peak-finding algorithms cannot be just blindly applied to the data to extract cosmic distance information, rather the opposite -- one should use cosmology-dependent fits of the full CF \cite{2009MNRAS.400.1643S, Sanchez:2012sg}. This would be a minor inconvenience if we knew how to predict the full non-linear galaxy CF as a function of only the cosmological parameters. Unfortunately, we are far from achieving that goal. 

In the past ten years, several approximate methods have been developed to extract cosmic distance information from BAO measurements. The most widely accepted technique defines the BAO scale in terms of a fiducial-model template CF, where the cosmological parameters are kept fixed at the fiducial values. {\it Ad hoc} nuisance parameters are added, to capture the effects of non-linearities and with the intent of ``marginalizing over'' the chosen fiducial cosmology \cite{2008ApJ...686...13S, 2012MNRAS.427.2146X}. This model template is then used to infer the cosmic distance from the statistical data analysis. Moreover, since non-linearities suppress the amplitude of the BAO, the observed galaxy positions are adjusted, using approximate non-linear model algorithms, to enhance the signal-to-noise of the BAO peak in the CF. This is done with the intent of restoring the pristine information on the acoustic scale; however, this {\it reconstruction} procedure explicitly depends on the choice of a fiducial cosmology and on the specification of a heuristic model of non-linear effects \cite{2007ApJ...664..675E}. Hence, in both  the treatment of the data and the statistical analysis, model-dependent assumptions intervene.  These carry the inherent risk of underestimating the uncertainties on cosmic distances, and potentially introduce a source of systematic bias in the cosmological-parameter inference. 

In order to overcome these limitations, a new promising BAO standard ruler in the galaxy CF, dubbed the {\it linear point} (LP hereafter), was suggested by some of us \cite{2016MNRAS.455.2474A}. Its position, defined as the midpoint between the positions of the peak and dip in the monopole CF, is located at $\sim95$ Mpc/h in comoving units \cite{2016MNRAS.455.2474A}. Using results from N-body simulations of $\Lambda$CDM models, it has been shown that the LP is insensitive to non-linear effects at $0.5\%$ relative to the linear-theory prediction. This holds for the matter-density field as well as for the spatial distribution of halos. Moreover, analytic arguments suggest that the LP remains stable (in both position and amplitude) with respect to the effects of redshift-space distortions and scale-dependent bias \cite{2016MNRAS.455.2474A}. An additional advantage of the LP is that it is a purely geometrical standard ruler, i.e.,  its position is independent of the amplitude and slope of the spectrum of primordial density fluctuations (at least for models similar to the $\Lambda$CDM scenario). Hence, unlike any other known BAO analysis, the LP can provide estimates of cosmic distances without the need for theoretical modelling of the CF data.

Recently, we have presented \cite{2017arXiv170301275A} a cosmological relation that allows us to infer the isotropic-volume distance $D_{V}$ using estimates of the LP from galaxy data. In particular, we focused on the CMASS and LOWZ galaxy samples from the Twelfth Data Release (DR12) of the Baryon Oscillation Spectroscopic Survey (BOSS)\footnote{\url{https://www.sdss3.org/surveys/boss.php}}, and found
\begin{eqnarray}
D_V^{\rm LP}(\bar{z}_{\rm LOWZ-DR12}=0.32) &=& (1264 \pm 28)\, {\rm Mpc}\, , \nn \\
D_V^{\rm LP}(\bar{z}_{\rm CMASS-DR12}=0.57) &=& (2056 \pm 22)\, {\rm Mpc}\, ,
\end{eqnarray}
thus providing distance estimates that are competitive with those obtained from standard assumption-rich BAO methods.

In this manuscript, we aim to validate the LP parametric model-independent estimation already applied in \cite{2017arXiv170301275A} to the actual LOWZ and CMASS  galaxy samples. To this end, we employ the Quick Particle Mesh (QPM) mock catalogues (``mocks'') \cite{2014MNRAS.437.2594W} built by the BOSS collaboration explicitly to mimic the LOWZ and CMASS clustering properties. They were largely used by the collaboration to test their Twelfth Data Release (DR12) BAO data analysis \cite{2016MNRAS.457.1770C}.

Our approach relies on a simple polynomial interpolation of the CF in the BAO range of scale. In this paper, we first validate the polynomial fit. Then, for each mock, the best-fit polynomial parameters and uncertainties provide the LP estimate and error. We find the optimal values of the polynomial order, the fitted range of scales, and the bin size to use for LP estimation on this data set. Optimization for future, larger-volume or higher-precision data sets would yield different values; but remarkably, our preliminary tests suggest that it will be sufficient just to shrink the fitted range of scales.

The paper is structured as follows. In Section \ref{sec:method}, we detail the methodology employed to validate the linear-point estimation through the polynomial fit: we summarize the characteristics of the QPM mocks, we define the systematic bias, we provide a checklist that the optimal fitting set-up should pass to be validated. In Section \ref{sec:LPposition}, we perform the previously introduced tests, discussing step by step the results of the analysis. In Section \ref{sec:concl}, we present our conclusions.


\section{Methodology}\label{sec:method}

In this section, we present the procedure developed to estimate the linear point from galaxy data. Our goal is to show that a simple model-independent parametric fit, applied to the monopole clustering correlation function, recovers the LP position without introducing systematic biases. We test this on mock catalogues, generated by the BOSS collaboration to reproduce the Luminous Red Galaxies (LRG) DR12-BAO clustering properties and used to test the BOSS BAO analysis \cite{2014MNRAS.437.2594W, 2016MNRAS.457.1770C}\footnote{Since we want to test the LP estimation procedure for different survey volumes we do not focus on the final galaxy clustering analysis performed by the BOSS collaboration \cite{2017MNRAS.470.2617A}. We consider instead the CF analysis presented in \cite{2016MNRAS.457.1770C} where the LOWZ and CMASS galaxy samples are taken into account.}

\subsection{QPM Mocks}
\label{sec:mocks}

QPM mocks \cite{2014MNRAS.437.2594W} were employed for the BOSS clustering analysis. The QPM method uses a low-resolution particle-mesh N-body solver. The halo catalogue and its properties were built to match the mass function and large-scale bias of halos of high-resolution simulations. The halo catalogue was then populated with galaxies using a halo occupation distribution (HOD) modelling, where the HOD parameters were adjusted for each mock by fitting the observed small-scale projected two-point galaxy correlation function for the LRGs. Each mock matches the angular and radial selection functions of the survey and the observed number density of galaxies. The final galaxy catalogue consists of 1000 realizations of the LOWZ sample and 956 for CMASS [17]. For each of these mocks, the CF has been computed using the Landy-Szalay algorithm \cite{1993ApJ...412...64L}.

The fiducial cosmology of the QPM mocks is a flat $\Lambda$CDM model, with cosmological parameter values close to the best-fit Planck+BOSS cosmology: $\Omega_{m} = 0.29$, $\Omega_{\Lambda}= 0.71$, $\Omega_{b}h^{2}=0.02247$, $\Omega_{\nu}h^{2}=0.0$, $h = 0.7$, $n_{s}=0.97$, and $\sigma_{8}=0.8$.

\subsection{Estimating the Linear Point position with a model-independent parametric fit}
\label{sec:extraction}

In order to extract the LP position from the galaxy monopole correlation function $\xi_{0}(s)$ ($s$ being the redshift-space coordinate in comoving units), we first estimate the positions of the maximum and the minimum of the CF in the BAO range of scales. This can readily be accomplished using a model-independent parametric fit. A simple, but (as we will see) efficient and robust, way to do so consists of first interpolating the CF data with a polynomial
\be
\label{poly}
\xi_{0}^{\rm fit}(s)=\sum_{i=0}^{n}a_i s^i\, ,
\ee 
where $n$ is the order of the polynomial fitting function. 
The solutions of $d\xi_{0}^{\rm fit}/ds=0$ are then computed, to find the location of the peak ($\hat{s}^{\rm fit}_{\rm peak}$) and dip ($\hat{s}^{\rm fit}_{\rm dip}$) in the CF.\footnote{It is worth noting that, in the analysis of real (rather than simulated) galaxy-survey data, one should account for the Alcock-Paczynski effect \cite{1979Natur.281..358A, 2013MNRAS.431.2834X}, which distorts the CF. In such a case, one can conveniently express the CF in terms of the dimensionless distance $y\equiv s/constant$ \cite{2016MNRAS.455.2474A, Sanchez:2012sg}. However, the procedure to extract the LP is the same whether the correlation function is expressed as function of $y$ or $s$. Therefore, to ease the reading of the present article, we work in comoving coordinates.} 
The estimated location of the LP is the mid-point between the computed dip and peak locations
\be
\label{LPestim}
\hat{s}^{\rm fit}_{LP}=\frac{1}{2}(\hat{s}^{\rm fit}_{\rm peak}+\hat{s}^{\rm fit}_{\rm dip})\, ,
\ee
which can be expressed in terms of the best-fit polynomial coefficients to the CF data.\footnote{
	To simplify the notation hereafter, we omit the hat and the ``fit'' subscripts.}
This allows us to estimate the uncertainty on the LP location, by propagating the uncertainties in the polynomial coefficients.

We would like to stress two considerations concerning the use of the polynomial interpolation of the CF. First, it provides an effective way of smoothing the noisy data points, thereby enabling the LP estimation. Indeed the more parameters we allow (i.e., the higher the order of the polynomial) the less effectively the polynomial fit smooths the CF. Nevertheless, we expect that the fitting procedure does not introduce a systematic bias in the determination of the LP, as we will show in Section \ref{D}. Second, the authors of \cite{2016MNRAS.455.2474A} found that, in the BAO range of scales, the CF is nearly anti-symmetric with respect to the LP. As we will show in Section \ref{E}, this provides us with a guideline to choose the optimal range of scales over which to interpolate the CF. 

In principle, the order of the CF polynomial-fitting function may depend on the range of scales considered, the redshift and the survey volume. Here, we find that an unbiased estimator of the LP requires $n\ge 5$. In the case of the LOWZ and CMASS mocks, we find that a quintic polynomial fit the CFs well over the range of scales considered. We will show this in Section \ref{D}, by comparing to the LP estimate obtained using a seventh-order polynomial.

\begin{figure}[h]
\centering
\includegraphics[width=0.9\hsize]{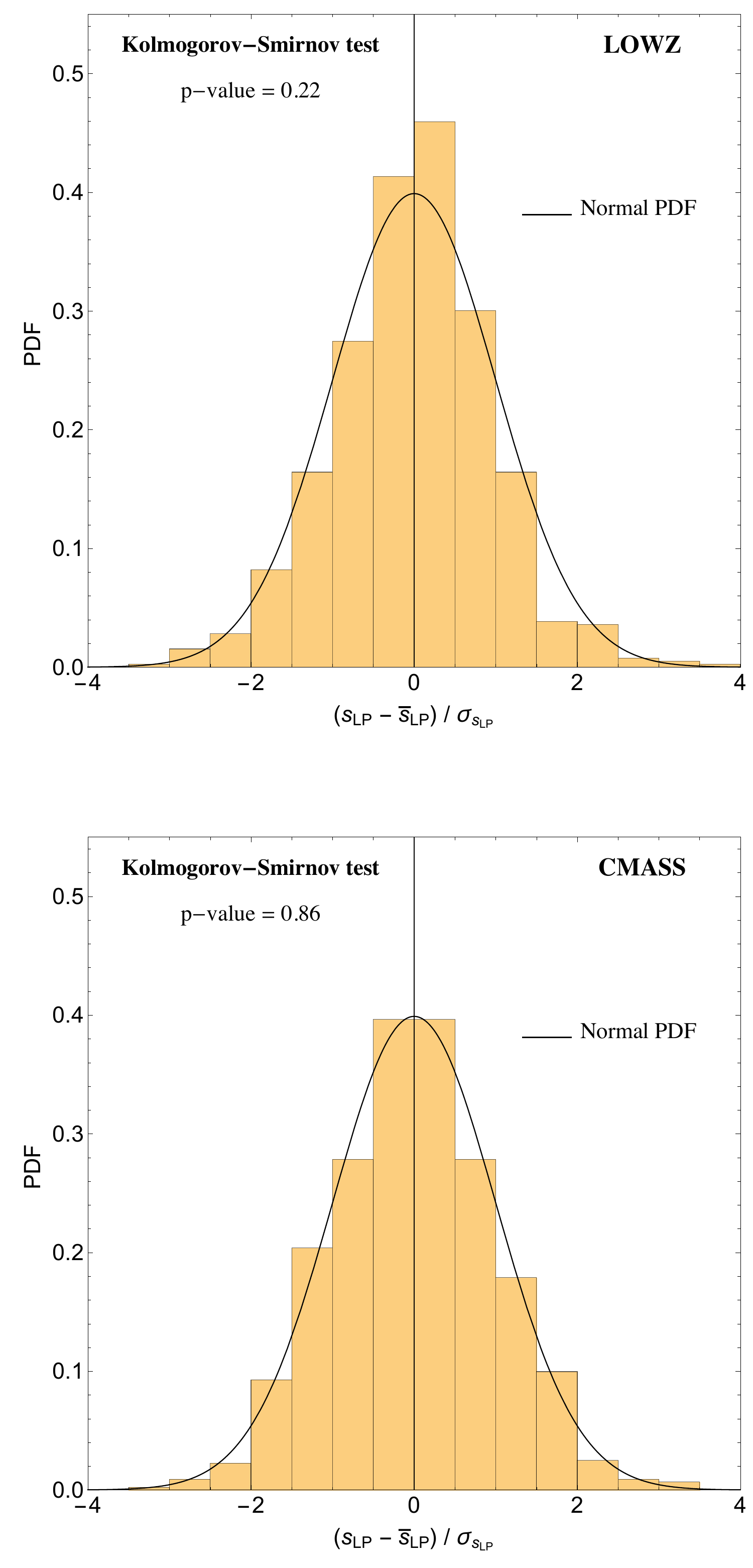}
\caption{\label{fig:KS} 
Normalized histograms of the rescaled linear-point positions recovered from the LOWZ {\it (upper panel)} and CMASS {\it (lower panel)} mock catalogues. The unit normal probability distribution function is overplotted. The $p$-values of the Kolmogorov-Smirnov test show that, for both catalogues, there is a reasonable probability that the LP values are drawn from a Gaussian distribution.}
\end{figure}

\subsection{Linear Point estimation: bias definition}
\label{sec:validation} 

Our analysis has two goals: on the one hand, we want to show that a simple polynomial fit to the CF can provide an unbiased estimate of the LP; on the other hand, we want to determine the optimal combination of polynomial order, range of scales, and binning that minimizes the LP statistical error. To this end, we introduce a measure of the LP systematic bias:
\be
\label{eq:bias}
b_{LP}=\bar{s}_{LP}-s_{LP}^{\rm true}\, .
\ee
Here, $\bar{s}_{LP}$ is the mean of the LP positions estimated
from the mocks, and $s_{LP}^{\rm true}$
is our reference LP value, which
we set to the value of the LP estimated from the average CF over the
mocks. This is because we are interested in evaluating 
only the uncertainty in the LP estimation due to the
polynomial-interpolation procedure, and not any small uncertainty that
arises due to the non-linear clustering 
of matter  \cite{2016MNRAS.455.2474A}. As already mentioned in Section I, that non-linear clustering systematically shifts the location of the
LP relative to the linear CF up to a $0.5\%$ \cite{2016MNRAS.455.2474A}, 
but this can  be mitigated by shifting the definition of the LP estimator relative to Eq. (\ref{LPestim}).
(This shift leaves Eq. (\ref{eq:bias}) unaffected, since it affects $\bar{s}_{LP}$ and $s_{LP}^{\rm true}$ equally.) 
We therefore take $s_{LP}^{\rm true}$ to be the value estimated from the CF
averaged over all the mocks. This has negligible cosmic-variance and sampling-variance error
compared to the individual mocks, but (deliberately) shares with them any potential systematic bias in $\bar{s}_{LP}$.\footnote{In \cite{2016MNRAS.455.2474A}, we have shown that the LP position in the CF, for both high-resolution N-body simulations and theoretical models, shifts with respect to  the linear theory prediction by no more than $1\%$. This shift is secular, and its effect is halved with a simple redshift-independent correction. In the case of the LOWZ and CMASS mocks, we find the ``true'' LP position to deviate by $1.3\%$ and $1.2\%$ w.r.t.~the linear theory prediction respectively. This disagreement could be due to the approximate treatment of clustering in the modeling used to built the QPM mocks \cite{2014MNRAS.437.2594W}.  Alternately,  it might be due to the way the HOD model is implemented in the QPM mocks. For instance, rather than adjusting the HOD parameters separately for each mock, one should properly fit the projected CF on small scales once for all the mocks (see e.g.~\cite{2017arXiv170804892S}). Furthermore, the parameter uncertainties due to the HOD fitting should be correctly propagated to the BAO scales. We plan to properly address these points with further investigations to be carried out with high-resolution and large-volume N-body simulations \cite{2014MNRAS.440.1420R} and improved implementations of the HOD \cite{2017arXiv170804892S}. Nevertheless, for the purpose of the present analysis, which is validating the non-linear LP extraction, we can safely ignore this issue.}
	
We would like to highlight that, given the finite volume of the mocks, the BAO feature in the CF might not be present in a given mock, due to cosmic variance. In this case, the polynomial estimator would not ``detect'' the peak and dip in the BAO range of scale, i.e.,  $d\xi_{0}^{\rm fit}/ds=0$ would have no actual solutions. Since the BAO is detected in the real data, we condition the LP error analysis to the mocks in which the LP is detected (for each polynomial-fit configuration). We therefore compute the conditional CF data covariance recursively: we perform a first polynomial fit of the CF of each mock using the covariance from the entire mock dataset, and, if the LP is not detected, we discard the mock and recompute the CF data covariance from the selected mocks. Depending on the polynomial fit configuration (order of the polynomial, bin size and range of scales), the fraction of retained mocks (Mock Acceptance Rate) is $\gtrsim 80\%$ for LOWZ and $\gtrsim 90\%$ for CMASS. As the statistical error on the LP position and bias are normalized to the number of retained mocks, the LP error analysis is independent of the specific value of the mock acceptance rate. Moreover, since there is no LP-position information in the mocks that do not show the BAO features, including them in the covariance-matrix computation does not change the LP error estimate. 

Given the finite size of the mock samples, we correct the LP error budget according to \cite{2007A&A...464..399H, 2014MNRAS.439.2531P}. Similarly, we follow \cite{2014MNRAS.439.2531P} for estimating errors on the determination of the fitting-polynomial coefficients for each mock's CF.

In summary, our validation of the LP estimation will assess the following points:
\begin{itemize}
  	\item[\textbf{A.}] Gaussianity of the correlation-function and linear point distributions: 
	we show that the both the CF and the LP are consistent with a Gaussian distribution.
  	\item[\textbf{B.}] Optimal polynomial estimator: 
	we consider polynomials of different orders as LP estimators, and discuss their suitability.
  	\item[\textbf{C.}] Optimal BAO range of scales: 
	we analyze the BAO range of scale fit for the CF, and identify the optimal one for LP estimation.
  	\item[\textbf{D.}] Optimal bin size: 
	we identify the bin sizes that return an unbiased LP estimate.
\end{itemize}

\section{Linear Point Estimation Tests}
\label{sec:LPposition} 

As mentioned in Section \ref{intro}, the advantage of the LP is that it is a geometric standard ruler on the BAO scale that is preserved by non-linear effects. LP estimation therefore does not require the use of reconstruction methods to be applied to galaxy catalogs. Hence, we test the LP estimation procedure on pre-reconstructed QPM mocks from the BOSS collaboration.

The results of the error evaluation, which will be presented below, indicate that the optimal setup for LP estimation consists of fitting the galaxy CF with a quintic polynomial estimator, in the range of scales $60< s$ [Mpc/h] $<130$, with bins of size $\Delta{s}=3$ Mpc/h.

\begin{table}
\vspace{1cm}
\renewcommand{\arraystretch}{1.1}
\begin{center}
\textbf{Estimator Test} \\
{\scriptsize
\begin{tabular}{l c c c c c c c c c c c c}\\
\hline 
\hline \\
Polyn. &&& $b_{LP}$   &&&  $\bar{\sigma}_{s_{LP}}$ &&& mean AIC$_{{\rm c}}$ \\ [0.3cm]
\hline 
\hline \\ 
\textbf{- LOWZ }  \\
Quintic &&& $-0.41$ Mpc/h  &&& $2.4$ Mpc/h  &&& 34    \\
7th order &&& $-0.37$ Mpc/h  &&& $2.7$ Mpc/h  &&& 41   \\ [0.6cm]
\textbf{- CMASS}\\
Quintic &&& $-0.25$ Mpc/h &&& $1.5$ Mpc/h &&& 35    \\
7th order &&& $-0.20$ Mpc/h &&& $1.7$ Mpc/h &&& 41    \\ [0.1cm]
\hline
\end{tabular}
}
\end{center}
\
\caption[]{\label{tab:est} 
We show the results of the {\it estimator test}. Both the quintic and the seventh-order polynomials are unbiased (i.e.,  negligible-bias) linear-point estimators. 
The quintic polynomial is the chosen LP estimator, as it provides the smallest errors and is preferred by the model-selection criterion.}
\end{table}

\begin{figure}[h]
\centering
\includegraphics[width=0.9\hsize]{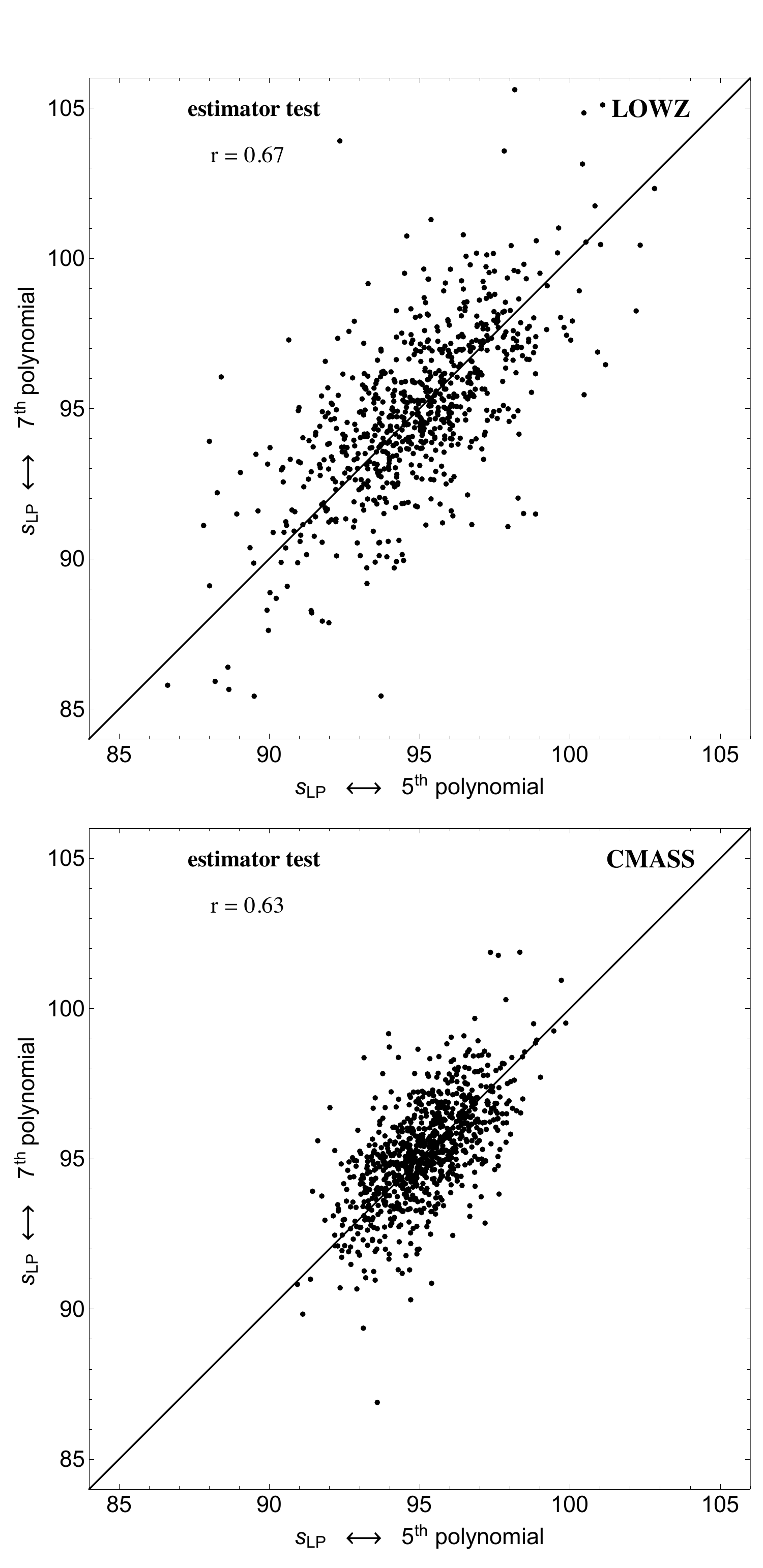}
\caption{\label{fig:estimator} 
Scatter in the LP {\it estimator} for the quintic {\it versus} the seventh-order polynomial fits to the CF of the LOWZ {\it (upper panel)} and CMASS {\it (lower panel)} mock catalogues. 
The scatter along the continuous black line indicates the cosmic-variance error, while the scatter perpendicular to the line represents the estimator error. 
The larger values of the correlation coefficients quoted in the panels suggest that the errors on the LP estimator are dominated by cosmic variance.
}
\end{figure}

\subsection{Gaussianity of the correlation function and linear point distributions} 
\label{B}

We verify that the distribution of the mock CF is always well-described by a Gaussian function. Hence, we can perform a statistical analysis of the CF assuming a Gaussian likelihood, and find the coefficients of a polynomial fitting function by simple $\chi^2$ minimization.

We check that the distribution of the $\chi^{2}_{{\rm min}}$ values from the polynomial fit to the CF of the mocks is consistent with a $\chi^{2}$ distribution, while the distribution of inferred values of $s_{LP}$ is consistent with a Gaussian. The latter is shown in Fig.~\ref{fig:KS}, where we plot the LOWZ {\it (upper panel)} and CMASS {\it (bottom panel)} normalized histograms of $(s_{LP}- \bar{s}_{LP})/ \sigma_{s_{LP}}$. The unit normal probability distribution function is overplotted. We perform the Kolmogorov-Smirnov test for the LOWZ and CMASS mocks. The $p$-values are respectively $0.22$ and $0.86$, indicating reasonable probabilities that the posterior of $s_{LP}$ is Gaussian distributed. Therefore, we can assign  the usual Gaussian meaning to the rms of the LP distribution.\footnote{
	We recall that, in this manuscript, we always use the error estimated from the likelihood and not from the distribution. However, after applying the corrections for the small number of mocks \cite{2014MNRAS.439.2531P}, the two agree to better than $7\%$.} 

The mean $s_{LP}$ errors for the two simulated galaxy samples are $\sigma^{{\rm LOWZ}}_{s_{LP}}=2.4$ Mpc/h and $\sigma^{{\rm CMASS}}_{s_{LP}}=1.5$ Mpc/h. Therefore, the intrinsic $0.5\%$ deviation of $s_{LP}$  with respect to $s^{{\rm lin}}_{LP}$ (found in \cite{2016MNRAS.455.2474A}) is subdominant.

\begin{figure}[h]
\centering
\includegraphics[width=0.9\hsize]{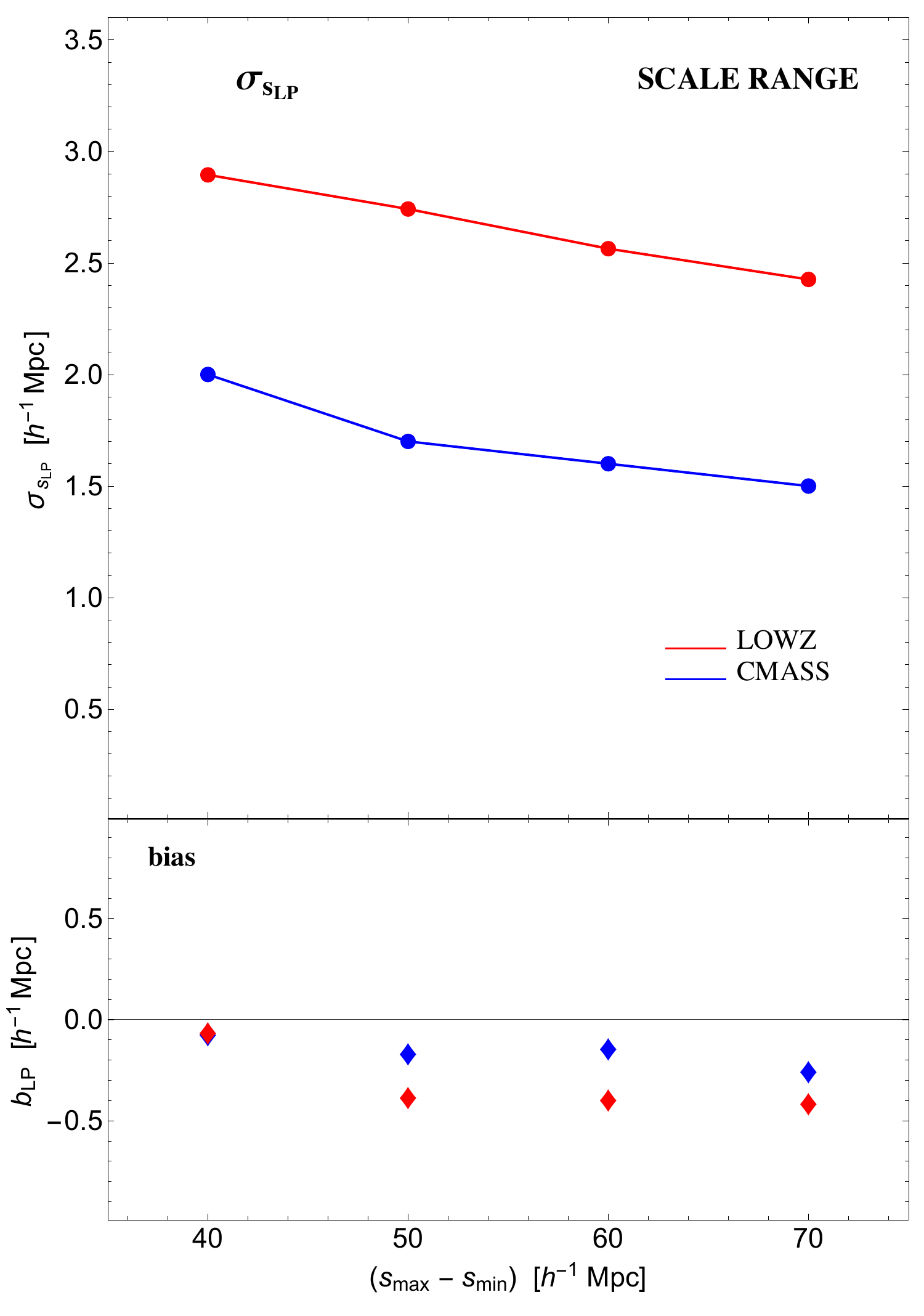}
\caption{\label{fig:ScaleError} 
LP-estimation error {\it (upper panel)} and bias {\it(lower panel)} as functions of the range of scales used to fit the CF. As we can see, the bias  is always negligible, since $b_{LP}\le 0.2 \times \bar{\sigma}_{s_{LP}}$.
}
\end{figure}

\subsection{Optimal polynomial estimator} 
\label{D}

To test the dependence of the LP estimation on the choice of the order of the polynomial fit to the CF, we consider a quintic polynomial and a seventh-order one. For each of these cases, we estimate $s_{LP}$ from Eq.~(\ref{LPestim}) and evaluate the bias as in Eq.~(\ref{eq:bias}). The results of the comparison are summarized in Table \ref{tab:est}, where we quote, for each mock, both the value of the bias and the average error on the LP estimator. In all cases, the absolute systematic shift is much smaller than the mean error estimated from the likelihood, i.e., $b_{LP}<0.2\times \bar{\sigma}_{LP}$. As expected, $\bar{\sigma}_{s^{{\rm 5th}}_{LP}}<\bar{\sigma}_{s^{{\rm 7th}}_{LP}}$, since the quintic-polynomial interpolation needs to fit a smaller number of parameters. 

In Fig. \ref{fig:estimator}, we show the scatter-plots of the recovered LP position for the two polynomial orders, using the LOWZ {\it (upper panel)} and CMASS {\it(lower panel)} mocks. The scatter along the solid line is an indication of the cosmic-variance error, while the scatter perpendicular to the solid line indicates the estimator error. Quantitatively, Pearson's correlation coefficient $r$ for the two samples, $r^{{\rm LOWZ}} = 0.67$ and $r^{{\rm CMASS}} = 0.63$, reveals that cosmic variance is the dominant error. We have checked that combined use of the two estimators does not significantly reduce the statistical error compared with using only the quintic-polynomial fit. Therefore, using only the quintic polynomial is sufficient for our purposes.

Since both of the estimators are unbiased, to choose between them we adopt a simple model-selection criterion: the finite-sample-corrected Akaike information criterion AIC$_{\text{c}}$ \citep{2007MNRAS.377L..74L}. We report its formula here for convenience (dropping an irrelevant additive constant):
\be
	\text{AIC}_{\text{c}} \equiv \chi^{2}_{\text{min}} + \frac{2 (n+1) N}{N-n-2}  \, ,
	\label{}
\ee
where $n$ was introduced in Eq.~(\ref{poly}) and $N$ is the number of  points fit.
The idea behind the AIC$_{\text{c}}$ is to balance the quality of fit to the observed data against the complexity of the model. The polynomial fit that gives the minimal AIC$_{\text{c}}$ value is  selected. From Table \ref{tab:est}, we see that the smallest mean AIC$_{\text{c}}$ belongs always to the quintic polynomial. This motivates its choice for the optimal set-up.

\subsection{Optimal BAO range of scales} 
\label{E}

We focus next on determining the optimal range of scales from which to extract the LP from measurements of the CF. 

The values of the coefficients of the best-fit polynomial to the CF, and their  associated errors, depend on the range of scale over which the CF is fit. This calls for selecting  an optimal range of scales that minimizes the statistical uncertainty, while introducing negligible systematic bias in the estimated LP location. We recall that the CF is anti-symmetric with respect to the linear point over the BAO range of scales \cite{2016MNRAS.455.2474A}. This motivates interpolating the CF with a quintic polynomial over a range that is symmetric with respect to $95$ Mpc/h.  

\begin{figure}[h]
\centering
\includegraphics[width=0.9\hsize]{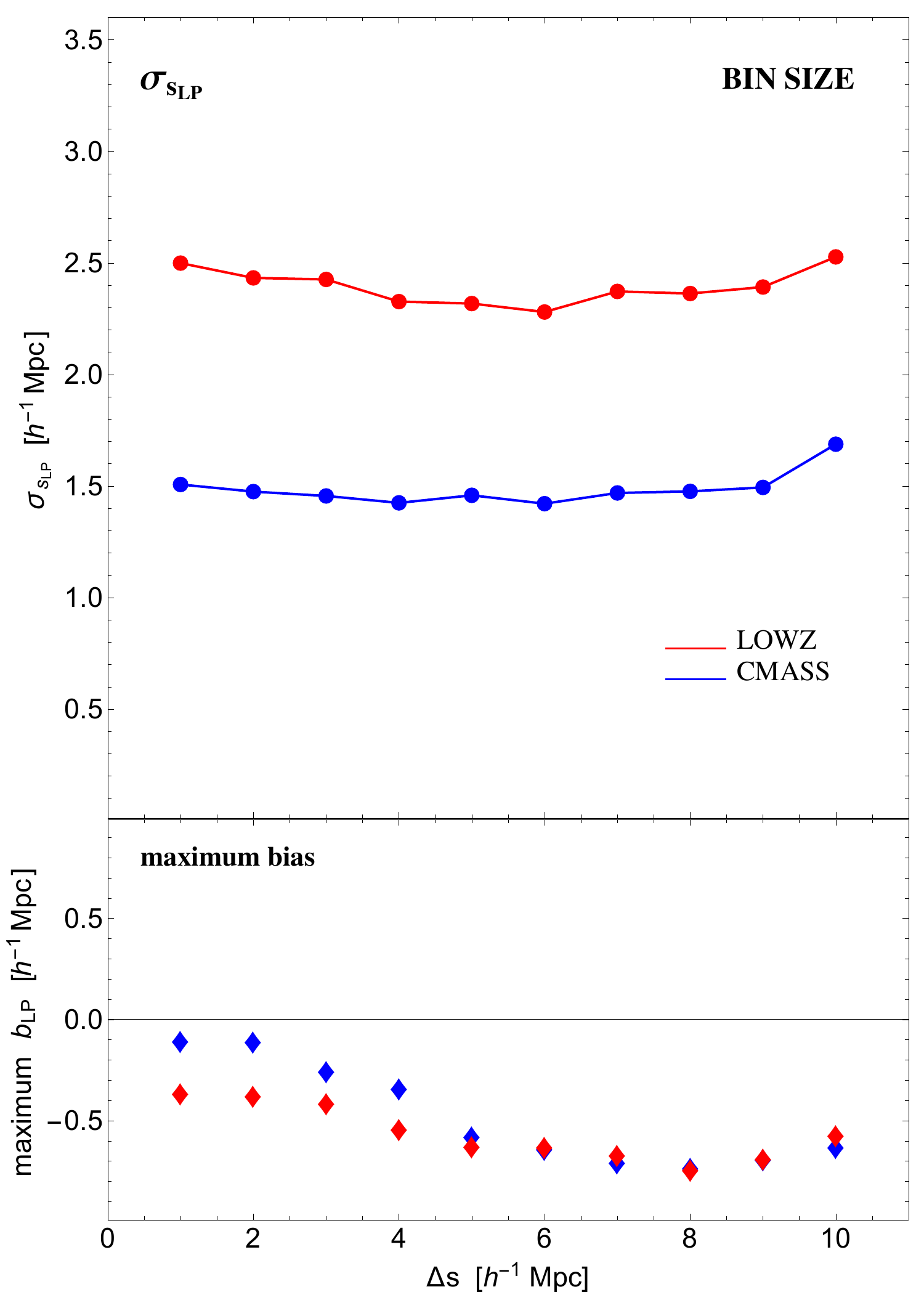}
\caption{\label{fig:bin_error} 
LP-estimation error  {\it (upper panel)} and bias {\it (lower panel)} as functions of the bin width. We see that the bias in the LP estimate induced by the choice of the binning of the CF is negligible (i.e., $b_{LP}\le 0.2 \times \bar{\sigma}_{s_{LP}}$) for $\Delta{s}\le 4$ Mpc/h.}
\end{figure}

In Fig.~\ref{fig:ScaleError}, we plot the statistical error {\it(upper panel)} and bias {\it(lower panel)} in the LP position, as functions of the interval of scales over which the CF is interpolated. We observe that, when the CF is interpolated over $(s_{{\rm max}}-s_{{\rm min}}) = 70$ Mpc/h, the statistical  error in the LP is minimized, while the systematic bias is negligible, $b_{LP} \leq 0.2 \times \bar{\sigma}_{s_{LP}}$. This trend is expected, as the fitting parameters are better determined when more information from the data is included.

We do not explore wider ranges of scale, since for the LOWZ real galaxy data,  due to the low signal-to-noise, the LP is detected only for $(s_{{\rm max}}-s_{{\rm min}}) \leq 70 $ Mpc/h  \cite{2017arXiv170301275A}. For the CMASS mock data, extending the fit over a larger range of scales does not result in a further reduction of the statistical errors. Thus we conclude that the optimal range of scales to is $(s_{\rm max}-s_{\rm min})=70$ Mpc/h.

\begin{figure}[h]
\centering
\includegraphics[width=0.9\hsize]{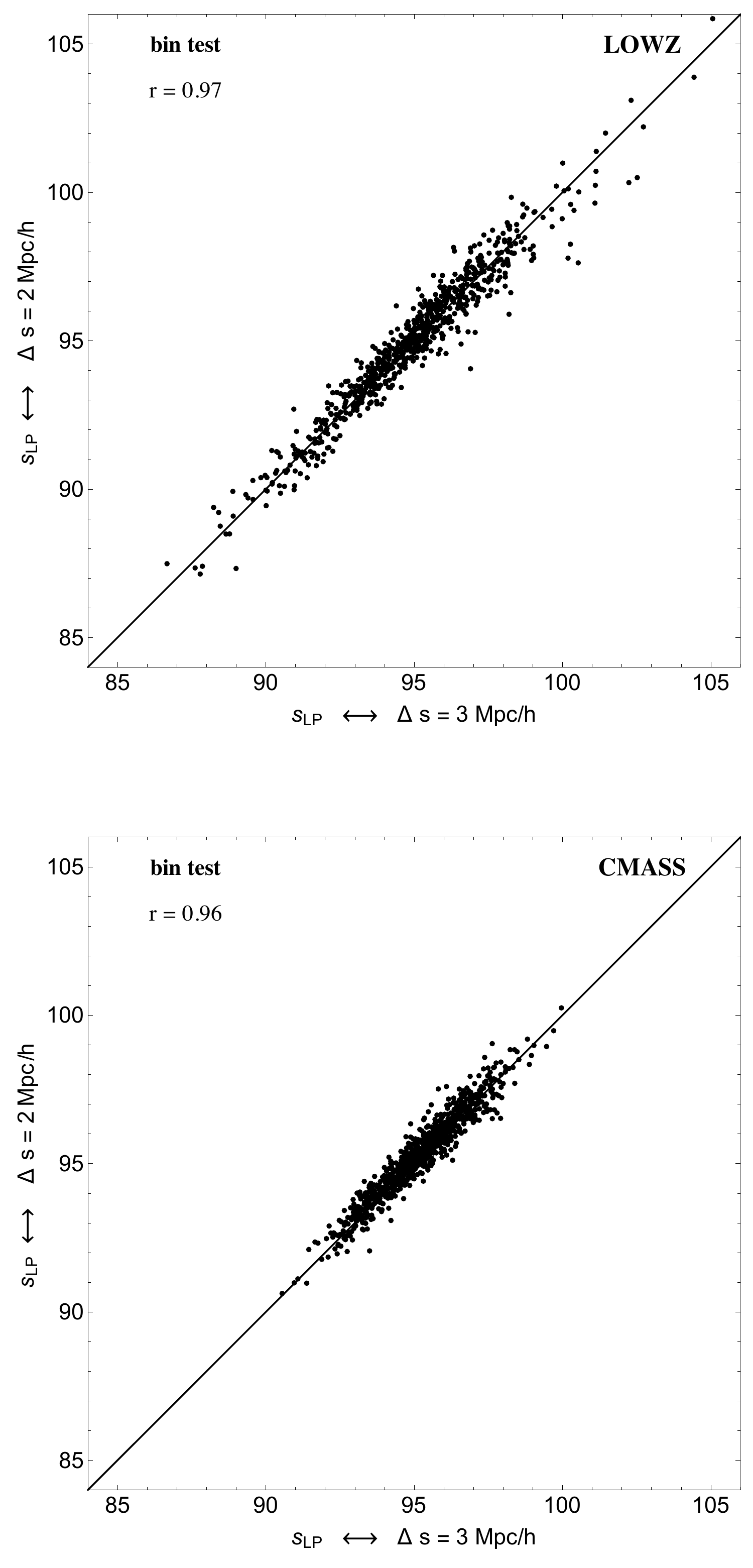}
\caption{\label{fig:bin_scatter} 
LP-estimator scatter plot, for a quintic-polynomial fit to the CF, with bins of width $\Delta{s}=2$ and $3$ Mpc/h, for the LOWZ {\it (upper panel)} and CMASS  {\it (lower panel)} mock catalogues.
}
\end{figure}

\subsection{Optimal bin size} 
\label{F}

The CF is measured in bins of finite width from datasets, whether simulated or real. In principle, the binning procedure can affect the LP estimation. To assess this effect, we have considered bins of varying size, from $\Delta{s}=1~\textrm{Mpc/h}$ to $\Delta{s}=10~\textrm{Mpc/h}$, and rebinned the CF mock data accordingly. In Fig. \ref{fig:bin_error}, we plot the values of $\sigma_{s_{LP}}$ and $b_{LP}$ corresponding to the most biased result among all the possible CF sampling possibilities for that bin size. While the LP statistical uncertainty is largely independent of the bin size, for $\Delta s \geq 5$ Mpc/h the mean LP value recovered from the mocks can be significantly biased. A too-large bin size introduces uncertainties in the bin positions, it does not provide enough sampling of the CF in the BAO range of scales, and it introduces a dependence on the sampling choice.

In Fig. \ref{fig:bin_scatter}, we show that the recovered LP position exhibits small scatter for small bin-sizes and, consequently, a high correlation coefficient $r$ between the $\Delta{s}=2~\textrm{Mpc/h}$ and the $\Delta{s}=3~\textrm{Mpc/h}$ LP estimators.

We conclude that, for $\Delta s \leq 4$ Mpc/h, the LP systematic bias is negligible; hence, recalling that a larger bin-size allows to reduce the covariance matrix noise \cite{2014MNRAS.439.2531P}, we choose $\Delta s = 3$ Mpc/h for the optimal set-up.

\section{Conclusions} 
\label{sec:concl}

Equipping the Baryon Acoustic Oscillations with a cosmological standard ruler is a highly desirable goal. It must be independent of the parameters characterizing the primordial fluctuations (within inflationary $\Lambda$CDM), and insensitive to non-linearities that develop during the late-time dark-energy-dominated era. The {\it linear point} provides such a ruler.

Another feature of the LP was not previously considered: its simple definition allows a model-independent BAO analysis. In this work, using mock galaxy catalogs, we have presented a validation of LP-estimation through a theory-free parametric fit to the galaxy CF. In \cite{2017arXiv170301275A}, we applied such an estimator to galaxy data, and showed that the method presented here holds, even when the Alcock-Paczynski distortion is present. We thus discovered that cosmological distances can be estimated without any need to model the non-linear physics that affects the galaxy correlation function at the BAO range of scales. 

In this paper, we have determined the optimal set-up to extract, by means of the LP, distance information from the BOSS-DR12 LOWZ and CMASS galaxy samples, justifying the methodology applied in \cite{2017arXiv170301275A}. This consists in using a quintic polynomial to fit the galaxy CF over the range of scales  $ 60 < s < 130 $ Mpc/h, with a bin width of $\Delta s = 3$ Mpc/h. 

We have noted that the LP observable is not detected in all the available mocks, because, for a small fraction of the mocks, the polynomial fits do not exhibit the peak-and-dip feature necessary for polynomial LP estimation. Since fits to both the LOWZ and CMASS data do exhibit the needed peak and dip \cite{2017arXiv170301275A}, we consistently condition our analysis to the mocks compatible with observations. It is important to stress that, as can be easily tested, this procedure does not change the LP error budget, since the excluded mocks do not carry information on the LP position. 


%

We plan to perform a LP-standard-ruler forecast analysis for future galaxy surveys such as Euclid (\url{http://sci.esa.int/euclid/}), DESI (\url{http://desi.lbl.gov}) and WFIRST (\url{https://wfirst.gsfc.nasa.gov}). 
The LP is also promising as a probe of the growth of structure, given that the amplitude of the CF at the LP is insensitive to non-linearities \cite{2016MNRAS.455.2474A}. Also worth investigating is the effect on the LP of massive neutrinos \cite{2015JCAP...01..034T, 2015JCAP...07..043C}, which is still not considered even in standard BAO analysis. The LP may also serve as a smoking gun of modified-gravity, especially if the BAO-LP anti-symmetric feature \cite{2016MNRAS.455.2474A} is altered in candidate models (such as Quasidilaton Massive Gravity Theory \cite{2015PhRvD..92h4033A, 2017PhRvD..96h4001A}). Alternatively, one could construct a maximal-deviation test where, in the context of the concordance $\Lambda$CDM, the maximal allowed deviation from the predicted CF anti-symmetry  feature (in the BAO regime) is quantified. 

\begin{acknowledgments} 
We are deeply grateful to Antonio Cuesta for providing us the correlation function QPM mock measurements, as well as for his careful reading of the manuscript and for his valuable suggestions. We thank Shadab Alam for discussions. The research leading to these results has received funding from the European Research Council under the European Community Seventh Framework Programme (FP7/2007-2013 Grant Agreement no. 279954) ERC-StG "EDECS". GDS is supported by a Department of Energy grant DE-SC0009946 to the particle astrophysics theory group at CWRU. IZ is supported by National Science Foundation grant AST-1612085.
\end{acknowledgments}

\bibliography{MyBib}

\begin{thebibliography}{32}
\expandafter\ifx\csname natexlab\endcsname\relax\def\natexlab#1{#1}\fi
\expandafter\ifx\csname bibnamefont\endcsname\relax
  \def\bibnamefont#1{#1}\fi
\expandafter\ifx\csname bibfnamefont\endcsname\relax
  \def\bibfnamefont#1{#1}\fi
\expandafter\ifx\csname citenamefont\endcsname\relax
  \def\citenamefont#1{#1}\fi
\expandafter\ifx\csname url\endcsname\relax
  \def\url#1{\texttt{#1}}\fi
\expandafter\ifx\csname urlprefix\endcsname\relax\def\urlprefix{URL }\fi
\providecommand{\bibinfo}[2]{#2}
\providecommand{\eprint}[2][]{\url{#2}}

\bibitem[{\citenamefont{{Anselmi}
  et~al.}(2017{\natexlab{a}})\citenamefont{{Anselmi}, {Starkman}, {Corasaniti},
  {Sheth}, and {Zehavi}}}]{2017arXiv170301275A}
\bibinfo{author}{\bibfnamefont{S.}~\bibnamefont{{Anselmi}}},
  \bibinfo{author}{\bibfnamefont{G.~D.} \bibnamefont{{Starkman}}},
  \bibinfo{author}{\bibfnamefont{P.-S.} \bibnamefont{{Corasaniti}}},
  \bibinfo{author}{\bibfnamefont{R.~K.} \bibnamefont{{Sheth}}},
  \bibnamefont{and} \bibinfo{author}{\bibfnamefont{I.}~\bibnamefont{{Zehavi}}},
  \bibinfo{journal}{ArXiv e-prints}  (\bibinfo{year}{2017}{\natexlab{a}}),
  \eprint{1703.01275}.

\bibitem[{\citenamefont{{Peebles} and {Yu}}(1970)}]{1970ApJ...162..815P}
\bibinfo{author}{\bibfnamefont{P.~J.~E.} \bibnamefont{{Peebles}}}
  \bibnamefont{and} \bibinfo{author}{\bibfnamefont{J.~T.} \bibnamefont{{Yu}}},
  \bibinfo{journal}{\apj} \textbf{\bibinfo{volume}{162}}, \bibinfo{pages}{815}
  (\bibinfo{year}{1970}).

\bibitem[{\citenamefont{{Sakharov}}(1966)}]{1966JETP...22..241S}
\bibinfo{author}{\bibfnamefont{A.~D.} \bibnamefont{{Sakharov}}},
  \bibinfo{journal}{Soviet Journal of Experimental and Theoretical Physics}
  \textbf{\bibinfo{volume}{22}}, \bibinfo{pages}{241} (\bibinfo{year}{1966}).

\bibitem[{\citenamefont{{Silk}}(1968)}]{1968ApJ...151..459S}
\bibinfo{author}{\bibfnamefont{J.}~\bibnamefont{{Silk}}},
  \bibinfo{journal}{\apj} \textbf{\bibinfo{volume}{151}}, \bibinfo{pages}{459}
  (\bibinfo{year}{1968}).

\bibitem[{\citenamefont{{Eisenstein} et~al.}(1998)\citenamefont{{Eisenstein},
  {Hu}, and {Tegmark}}}]{1998ApJ...504L..57E}
\bibinfo{author}{\bibfnamefont{D.~J.} \bibnamefont{{Eisenstein}}},
  \bibinfo{author}{\bibfnamefont{W.}~\bibnamefont{{Hu}}}, \bibnamefont{and}
  \bibinfo{author}{\bibfnamefont{M.}~\bibnamefont{{Tegmark}}},
  \bibinfo{journal}{\apjl} \textbf{\bibinfo{volume}{504}}, \bibinfo{pages}{L57}
  (\bibinfo{year}{1998}), \eprint{astro-ph/9805239}.

\bibitem[{\citenamefont{Bassett and Hlozek}(2009)}]{Bassett:2009mm}
\bibinfo{author}{\bibfnamefont{B.~A.} \bibnamefont{Bassett}} \bibnamefont{and}
  \bibinfo{author}{\bibfnamefont{R.}~\bibnamefont{Hlozek}}
  (\bibinfo{year}{2009}), \eprint{0910.5224}.

\bibitem[{\citenamefont{{Eisenstein} et~al.}(2005)\citenamefont{{Eisenstein},
  {Zehavi}, {Hogg}, {Scoccimarro}, {Blanton}, {Nichol}, {Scranton}, {Seo},
  {Tegmark}, {Zheng} et~al.}}]{2005ApJ...633..560E}
\bibinfo{author}{\bibfnamefont{D.~J.} \bibnamefont{{Eisenstein}}},
  \bibinfo{author}{\bibfnamefont{I.}~\bibnamefont{{Zehavi}}},
  \bibinfo{author}{\bibfnamefont{D.~W.} \bibnamefont{{Hogg}}},
  \bibinfo{author}{\bibfnamefont{R.}~\bibnamefont{{Scoccimarro}}},
  \bibinfo{author}{\bibfnamefont{M.~R.} \bibnamefont{{Blanton}}},
  \bibinfo{author}{\bibfnamefont{R.~C.} \bibnamefont{{Nichol}}},
  \bibinfo{author}{\bibfnamefont{R.}~\bibnamefont{{Scranton}}},
  \bibinfo{author}{\bibfnamefont{H.-J.} \bibnamefont{{Seo}}},
  \bibinfo{author}{\bibfnamefont{M.}~\bibnamefont{{Tegmark}}},
  \bibinfo{author}{\bibfnamefont{Z.}~\bibnamefont{{Zheng}}},
  \bibnamefont{et~al.}, \bibinfo{journal}{\apj} \textbf{\bibinfo{volume}{633}},
  \bibinfo{pages}{560} (\bibinfo{year}{2005}), \eprint{astro-ph/0501171}.

\bibitem[{\citenamefont{{Smith} et~al.}(2008)\citenamefont{{Smith},
  {Scoccimarro}, and {Sheth}}}]{2008PhRvD..77d3525S}
\bibinfo{author}{\bibfnamefont{R.~E.} \bibnamefont{{Smith}}},
  \bibinfo{author}{\bibfnamefont{R.}~\bibnamefont{{Scoccimarro}}},
  \bibnamefont{and} \bibinfo{author}{\bibfnamefont{R.~K.}
  \bibnamefont{{Sheth}}}, \bibinfo{journal}{Phys. Rev. D}
  \textbf{\bibinfo{volume}{77}}, \bibinfo{eid}{043525} (\bibinfo{year}{2008}),
  \eprint{astro-ph/0703620}.

\bibitem[{\citenamefont{{Eisenstein}
  et~al.}(2007{\natexlab{a}})\citenamefont{{Eisenstein}, {Seo}, and
  {White}}}]{2007ApJ...664..660E}
\bibinfo{author}{\bibfnamefont{D.~J.} \bibnamefont{{Eisenstein}}},
  \bibinfo{author}{\bibfnamefont{H.-J.} \bibnamefont{{Seo}}}, \bibnamefont{and}
  \bibinfo{author}{\bibfnamefont{M.}~\bibnamefont{{White}}},
  \bibinfo{journal}{\apj} \textbf{\bibinfo{volume}{664}}, \bibinfo{pages}{660}
  (\bibinfo{year}{2007}{\natexlab{a}}), \eprint{astro-ph/0604361}.

\bibitem[{\citenamefont{Matsubara}(2008)}]{Matsubara:2007wj}
\bibinfo{author}{\bibfnamefont{T.}~\bibnamefont{Matsubara}},
  \bibinfo{journal}{Phys.Rev.} \textbf{\bibinfo{volume}{D77}},
  \bibinfo{pages}{063530} (\bibinfo{year}{2008}), \eprint{0711.2521}.

\bibitem[{\citenamefont{Crocce and Scoccimarro}(2008)}]{Crocce:2007dt}
\bibinfo{author}{\bibfnamefont{M.}~\bibnamefont{Crocce}} \bibnamefont{and}
  \bibinfo{author}{\bibfnamefont{R.}~\bibnamefont{Scoccimarro}},
  \bibinfo{journal}{Phys. Rev.} \textbf{\bibinfo{volume}{D77}},
  \bibinfo{pages}{023533} (\bibinfo{year}{2008}), \eprint{0704.2783}.

\bibitem[{\citenamefont{{S{\'a}nchez} et~al.}(2009)\citenamefont{{S{\'a}nchez},
  {Crocce}, {Cabr{\'e}}, {Baugh}, and {Gazta{\~n}aga}}}]{2009MNRAS.400.1643S}
\bibinfo{author}{\bibfnamefont{A.~G.} \bibnamefont{{S{\'a}nchez}}},
  \bibinfo{author}{\bibfnamefont{M.}~\bibnamefont{{Crocce}}},
  \bibinfo{author}{\bibfnamefont{A.}~\bibnamefont{{Cabr{\'e}}}},
  \bibinfo{author}{\bibfnamefont{C.~M.} \bibnamefont{{Baugh}}},
  \bibnamefont{and}
  \bibinfo{author}{\bibfnamefont{E.}~\bibnamefont{{Gazta{\~n}aga}}},
  \bibinfo{journal}{\mnras} \textbf{\bibinfo{volume}{400}},
  \bibinfo{pages}{1643} (\bibinfo{year}{2009}), \eprint{0901.2570}.

\bibitem[{\citenamefont{Sanchez et~al.}(2012)\citenamefont{Sanchez, Scoccola,
  Ross, Percival, Manera et~al.}}]{Sanchez:2012sg}
\bibinfo{author}{\bibfnamefont{A.~G.} \bibnamefont{Sanchez}},
  \bibinfo{author}{\bibfnamefont{C.}~\bibnamefont{Scoccola}},
  \bibinfo{author}{\bibfnamefont{A.}~\bibnamefont{Ross}},
  \bibinfo{author}{\bibfnamefont{W.}~\bibnamefont{Percival}},
  \bibinfo{author}{\bibfnamefont{M.}~\bibnamefont{Manera}},
  \bibnamefont{et~al.}, \bibinfo{journal}{Mon.Not.Roy.Astron.Soc.}
  \textbf{\bibinfo{volume}{425}}, \bibinfo{pages}{415} (\bibinfo{year}{2012}),
  \eprint{1203.6616}.

\bibitem[{\citenamefont{{Seo} et~al.}(2008)\citenamefont{{Seo}, {Siegel},
  {Eisenstein}, and {White}}}]{2008ApJ...686...13S}
\bibinfo{author}{\bibfnamefont{H.-J.} \bibnamefont{{Seo}}},
  \bibinfo{author}{\bibfnamefont{E.~R.} \bibnamefont{{Siegel}}},
  \bibinfo{author}{\bibfnamefont{D.~J.} \bibnamefont{{Eisenstein}}},
  \bibnamefont{and} \bibinfo{author}{\bibfnamefont{M.}~\bibnamefont{{White}}},
  \bibinfo{journal}{\apj} \textbf{\bibinfo{volume}{686}}, \bibinfo{pages}{13}
  (\bibinfo{year}{2008}), \eprint{0805.0117}.

\bibitem[{\citenamefont{{Xu} et~al.}(2012)\citenamefont{{Xu}, {Padmanabhan},
  {Eisenstein}, {Mehta}, and {Cuesta}}}]{2012MNRAS.427.2146X}
\bibinfo{author}{\bibfnamefont{X.}~\bibnamefont{{Xu}}},
  \bibinfo{author}{\bibfnamefont{N.}~\bibnamefont{{Padmanabhan}}},
  \bibinfo{author}{\bibfnamefont{D.~J.} \bibnamefont{{Eisenstein}}},
  \bibinfo{author}{\bibfnamefont{K.~T.} \bibnamefont{{Mehta}}},
  \bibnamefont{and} \bibinfo{author}{\bibfnamefont{A.~J.}
  \bibnamefont{{Cuesta}}}, \bibinfo{journal}{\mnras}
  \textbf{\bibinfo{volume}{427}}, \bibinfo{pages}{2146} (\bibinfo{year}{2012}),
  \eprint{1202.0091}.

\bibitem[{\citenamefont{{Eisenstein}
  et~al.}(2007{\natexlab{b}})\citenamefont{{Eisenstein}, {Seo}, {Sirko}, and
  {Spergel}}}]{2007ApJ...664..675E}
\bibinfo{author}{\bibfnamefont{D.~J.} \bibnamefont{{Eisenstein}}},
  \bibinfo{author}{\bibfnamefont{H.-J.} \bibnamefont{{Seo}}},
  \bibinfo{author}{\bibfnamefont{E.}~\bibnamefont{{Sirko}}}, \bibnamefont{and}
  \bibinfo{author}{\bibfnamefont{D.~N.} \bibnamefont{{Spergel}}},
  \bibinfo{journal}{\apj} \textbf{\bibinfo{volume}{664}}, \bibinfo{pages}{675}
  (\bibinfo{year}{2007}{\natexlab{b}}), \eprint{astro-ph/0604362}.

\bibitem[{\citenamefont{{Anselmi} et~al.}(2016)\citenamefont{{Anselmi},
  {Starkman}, and {Sheth}}}]{2016MNRAS.455.2474A}
\bibinfo{author}{\bibfnamefont{S.}~\bibnamefont{{Anselmi}}},
  \bibinfo{author}{\bibfnamefont{G.~D.} \bibnamefont{{Starkman}}},
  \bibnamefont{and} \bibinfo{author}{\bibfnamefont{R.~K.}
  \bibnamefont{{Sheth}}}, \bibinfo{journal}{\mnras}
  \textbf{\bibinfo{volume}{455}}, \bibinfo{pages}{2474} (\bibinfo{year}{2016}),
  \eprint{1508.01170}.

\bibitem[{\citenamefont{{White} et~al.}(2014)\citenamefont{{White}, {Tinker},
  and {McBride}}}]{2014MNRAS.437.2594W}
\bibinfo{author}{\bibfnamefont{M.}~\bibnamefont{{White}}},
  \bibinfo{author}{\bibfnamefont{J.~L.} \bibnamefont{{Tinker}}},
  \bibnamefont{and} \bibinfo{author}{\bibfnamefont{C.~K.}
  \bibnamefont{{McBride}}}, \bibinfo{journal}{\mnras}
  \textbf{\bibinfo{volume}{437}}, \bibinfo{pages}{2594} (\bibinfo{year}{2014}),
  \eprint{1309.5532}.

\bibitem[{\citenamefont{{Cuesta} et~al.}(2016)\citenamefont{{Cuesta},
  {Vargas-Maga{\~n}a}, {Beutler}, {Bolton}, {Brownstein}, {Eisenstein},
  {Gil-Mar{\'{\i}}n}, {Ho}, {McBride}, {Maraston}
  et~al.}}]{2016MNRAS.457.1770C}
\bibinfo{author}{\bibfnamefont{A.~J.} \bibnamefont{{Cuesta}}},
  \bibinfo{author}{\bibfnamefont{M.}~\bibnamefont{{Vargas-Maga{\~n}a}}},
  \bibinfo{author}{\bibfnamefont{F.}~\bibnamefont{{Beutler}}},
  \bibinfo{author}{\bibfnamefont{A.~S.} \bibnamefont{{Bolton}}},
  \bibinfo{author}{\bibfnamefont{J.~R.} \bibnamefont{{Brownstein}}},
  \bibinfo{author}{\bibfnamefont{D.~J.} \bibnamefont{{Eisenstein}}},
  \bibinfo{author}{\bibfnamefont{H.}~\bibnamefont{{Gil-Mar{\'{\i}}n}}},
  \bibinfo{author}{\bibfnamefont{S.}~\bibnamefont{{Ho}}},
  \bibinfo{author}{\bibfnamefont{C.~K.} \bibnamefont{{McBride}}},
  \bibinfo{author}{\bibfnamefont{C.}~\bibnamefont{{Maraston}}},
  \bibnamefont{et~al.}, \bibinfo{journal}{\mnras}
  \textbf{\bibinfo{volume}{457}}, \bibinfo{pages}{1770} (\bibinfo{year}{2016}),
  \eprint{1509.06371}.

\bibitem[{\citenamefont{{Alam} et~al.}(2017)\citenamefont{{Alam}, {Ata},
  {Bailey}, {Beutler}, {Bizyaev}, {Blazek}, {Bolton}, {Brownstein}, {Burden},
  {Chuang} et~al.}}]{2017MNRAS.470.2617A}
\bibinfo{author}{\bibfnamefont{S.}~\bibnamefont{{Alam}}},
  \bibinfo{author}{\bibfnamefont{M.}~\bibnamefont{{Ata}}},
  \bibinfo{author}{\bibfnamefont{S.}~\bibnamefont{{Bailey}}},
  \bibinfo{author}{\bibfnamefont{F.}~\bibnamefont{{Beutler}}},
  \bibinfo{author}{\bibfnamefont{D.}~\bibnamefont{{Bizyaev}}},
  \bibinfo{author}{\bibfnamefont{J.~A.} \bibnamefont{{Blazek}}},
  \bibinfo{author}{\bibfnamefont{A.~S.} \bibnamefont{{Bolton}}},
  \bibinfo{author}{\bibfnamefont{J.~R.} \bibnamefont{{Brownstein}}},
  \bibinfo{author}{\bibfnamefont{A.}~\bibnamefont{{Burden}}},
  \bibinfo{author}{\bibfnamefont{C.-H.} \bibnamefont{{Chuang}}},
  \bibnamefont{et~al.}, \bibinfo{journal}{\mnras}
  \textbf{\bibinfo{volume}{470}}, \bibinfo{pages}{2617} (\bibinfo{year}{2017}),
  \eprint{1607.03155}.

\bibitem[{\citenamefont{{Landy} and {Szalay}}(1993)}]{1993ApJ...412...64L}
\bibinfo{author}{\bibfnamefont{S.~D.} \bibnamefont{{Landy}}} \bibnamefont{and}
  \bibinfo{author}{\bibfnamefont{A.~S.} \bibnamefont{{Szalay}}},
  \bibinfo{journal}{\apj} \textbf{\bibinfo{volume}{412}}, \bibinfo{pages}{64}
  (\bibinfo{year}{1993}).

\bibitem[{\citenamefont{{Alcock} and {Paczynski}}(1979)}]{1979Natur.281..358A}
\bibinfo{author}{\bibfnamefont{C.}~\bibnamefont{{Alcock}}} \bibnamefont{and}
  \bibinfo{author}{\bibfnamefont{B.}~\bibnamefont{{Paczynski}}},
  \bibinfo{journal}{\nat} \textbf{\bibinfo{volume}{281}}, \bibinfo{pages}{358}
  (\bibinfo{year}{1979}).

\bibitem[{\citenamefont{{Xu} et~al.}(2013)\citenamefont{{Xu}, {Cuesta},
  {Padmanabhan}, {Eisenstein}, and {McBride}}}]{2013MNRAS.431.2834X}
\bibinfo{author}{\bibfnamefont{X.}~\bibnamefont{{Xu}}},
  \bibinfo{author}{\bibfnamefont{A.~J.} \bibnamefont{{Cuesta}}},
  \bibinfo{author}{\bibfnamefont{N.}~\bibnamefont{{Padmanabhan}}},
  \bibinfo{author}{\bibfnamefont{D.~J.} \bibnamefont{{Eisenstein}}},
  \bibnamefont{and} \bibinfo{author}{\bibfnamefont{C.~K.}
  \bibnamefont{{McBride}}}, \bibinfo{journal}{\mnras}
  \textbf{\bibinfo{volume}{431}}, \bibinfo{pages}{2834} (\bibinfo{year}{2013}),
  \eprint{1206.6732}.

\bibitem[{\citenamefont{{Sinha} et~al.}(2017)\citenamefont{{Sinha}, {Berlind},
  {McBride}, {Scoccimarro}, {Piscionere}, and {Wibking}}}]{2017arXiv170804892S}
\bibinfo{author}{\bibfnamefont{M.}~\bibnamefont{{Sinha}}},
  \bibinfo{author}{\bibfnamefont{A.~A.} \bibnamefont{{Berlind}}},
  \bibinfo{author}{\bibfnamefont{C.~K.} \bibnamefont{{McBride}}},
  \bibinfo{author}{\bibfnamefont{R.}~\bibnamefont{{Scoccimarro}}},
  \bibinfo{author}{\bibfnamefont{J.~A.} \bibnamefont{{Piscionere}}},
  \bibnamefont{and} \bibinfo{author}{\bibfnamefont{B.~D.}
  \bibnamefont{{Wibking}}}, \bibinfo{journal}{ArXiv e-prints}
  (\bibinfo{year}{2017}), \eprint{1708.04892}.

\bibitem[{\citenamefont{{Rasera} et~al.}(2014)\citenamefont{{Rasera},
  {Corasaniti}, {Alimi}, {Bouillot}, {Reverdy}, and
  {Balm{\`e}s}}}]{2014MNRAS.440.1420R}
\bibinfo{author}{\bibfnamefont{Y.}~\bibnamefont{{Rasera}}},
  \bibinfo{author}{\bibfnamefont{P.-S.} \bibnamefont{{Corasaniti}}},
  \bibinfo{author}{\bibfnamefont{J.-M.} \bibnamefont{{Alimi}}},
  \bibinfo{author}{\bibfnamefont{V.}~\bibnamefont{{Bouillot}}},
  \bibinfo{author}{\bibfnamefont{V.}~\bibnamefont{{Reverdy}}},
  \bibnamefont{and}
  \bibinfo{author}{\bibfnamefont{I.}~\bibnamefont{{Balm{\`e}s}}},
  \bibinfo{journal}{\mnras} \textbf{\bibinfo{volume}{440}},
  \bibinfo{pages}{1420} (\bibinfo{year}{2014}), \eprint{1311.5662}.

\bibitem[{\citenamefont{{Hartlap} et~al.}(2007)\citenamefont{{Hartlap},
  {Simon}, and {Schneider}}}]{2007A&A...464..399H}
\bibinfo{author}{\bibfnamefont{J.}~\bibnamefont{{Hartlap}}},
  \bibinfo{author}{\bibfnamefont{P.}~\bibnamefont{{Simon}}}, \bibnamefont{and}
  \bibinfo{author}{\bibfnamefont{P.}~\bibnamefont{{Schneider}}},
  \bibinfo{journal}{\aap} \textbf{\bibinfo{volume}{464}}, \bibinfo{pages}{399}
  (\bibinfo{year}{2007}), \eprint{astro-ph/0608064}.

\bibitem[{\citenamefont{{Percival} et~al.}(2014)\citenamefont{{Percival},
  {Ross}, {S{\'a}nchez}, {Samushia}, {Burden}, {Crittenden}, {Cuesta},
  {Magana}, {Manera}, {Beutler} et~al.}}]{2014MNRAS.439.2531P}
\bibinfo{author}{\bibfnamefont{W.~J.} \bibnamefont{{Percival}}},
  \bibinfo{author}{\bibfnamefont{A.~J.} \bibnamefont{{Ross}}},
  \bibinfo{author}{\bibfnamefont{A.~G.} \bibnamefont{{S{\'a}nchez}}},
  \bibinfo{author}{\bibfnamefont{L.}~\bibnamefont{{Samushia}}},
  \bibinfo{author}{\bibfnamefont{A.}~\bibnamefont{{Burden}}},
  \bibinfo{author}{\bibfnamefont{R.}~\bibnamefont{{Crittenden}}},
  \bibinfo{author}{\bibfnamefont{A.~J.} \bibnamefont{{Cuesta}}},
  \bibinfo{author}{\bibfnamefont{M.~V.} \bibnamefont{{Magana}}},
  \bibinfo{author}{\bibfnamefont{M.}~\bibnamefont{{Manera}}},
  \bibinfo{author}{\bibfnamefont{F.}~\bibnamefont{{Beutler}}},
  \bibnamefont{et~al.}, \bibinfo{journal}{\mnras}
  \textbf{\bibinfo{volume}{439}}, \bibinfo{pages}{2531} (\bibinfo{year}{2014}),
  \eprint{1312.4841}.

\bibitem[{\citenamefont{{Liddle}}(2007)}]{2007MNRAS.377L..74L}
\bibinfo{author}{\bibfnamefont{A.~R.} \bibnamefont{{Liddle}}},
  \bibinfo{journal}{\mnras} \textbf{\bibinfo{volume}{377}},
  \bibinfo{pages}{L74} (\bibinfo{year}{2007}), \eprint{astro-ph/0701113}.

\bibitem[{\citenamefont{{Thepsuriya} and {Lewis}}(2015)}]{2015JCAP...01..034T}
\bibinfo{author}{\bibfnamefont{K.}~\bibnamefont{{Thepsuriya}}}
  \bibnamefont{and} \bibinfo{author}{\bibfnamefont{A.}~\bibnamefont{{Lewis}}},
  \bibinfo{journal}{\jcap} \textbf{\bibinfo{volume}{1}}, \bibinfo{eid}{034}
  (\bibinfo{year}{2015}), \eprint{1409.5066}.

\bibitem[{\citenamefont{{Castorina} et~al.}(2015)\citenamefont{{Castorina},
  {Carbone}, {Bel}, {Sefusatti}, and {Dolag}}}]{2015JCAP...07..043C}
\bibinfo{author}{\bibfnamefont{E.}~\bibnamefont{{Castorina}}},
  \bibinfo{author}{\bibfnamefont{C.}~\bibnamefont{{Carbone}}},
  \bibinfo{author}{\bibfnamefont{J.}~\bibnamefont{{Bel}}},
  \bibinfo{author}{\bibfnamefont{E.}~\bibnamefont{{Sefusatti}}},
  \bibnamefont{and} \bibinfo{author}{\bibfnamefont{K.}~\bibnamefont{{Dolag}}},
  \bibinfo{journal}{\jcap} \textbf{\bibinfo{volume}{7}}, \bibinfo{eid}{043}
  (\bibinfo{year}{2015}), \eprint{1505.07148}.

\bibitem[{\citenamefont{{Anselmi} et~al.}(2015)\citenamefont{{Anselmi},
  {L{\'o}pez Nacir}, and {Starkman}}}]{2015PhRvD..92h4033A}
\bibinfo{author}{\bibfnamefont{S.}~\bibnamefont{{Anselmi}}},
  \bibinfo{author}{\bibfnamefont{D.}~\bibnamefont{{L{\'o}pez Nacir}}},
  \bibnamefont{and} \bibinfo{author}{\bibfnamefont{G.~D.}
  \bibnamefont{{Starkman}}}, \bibinfo{journal}{\prd}
  \textbf{\bibinfo{volume}{92}}, \bibinfo{eid}{084033} (\bibinfo{year}{2015}),
  \eprint{1506.01000}.

\bibitem[{\citenamefont{{Anselmi}
  et~al.}(2017{\natexlab{b}})\citenamefont{{Anselmi}, {Kumar}, {L{\'o}pez
  Nacir}, and {Starkman}}}]{2017PhRvD..96h4001A}
\bibinfo{author}{\bibfnamefont{S.}~\bibnamefont{{Anselmi}}},
  \bibinfo{author}{\bibfnamefont{S.}~\bibnamefont{{Kumar}}},
  \bibinfo{author}{\bibfnamefont{D.}~\bibnamefont{{L{\'o}pez Nacir}}},
  \bibnamefont{and} \bibinfo{author}{\bibfnamefont{G.~D.}
  \bibnamefont{{Starkman}}}, \bibinfo{journal}{\prd}
  \textbf{\bibinfo{volume}{96}}, \bibinfo{eid}{084001}
  (\bibinfo{year}{2017}{\natexlab{b}}), \eprint{1706.01872}.

\end{thebibliography}

\end{document}